\documentstyle[12pt,epsf]{article}
\advance\textheight by \topskip
\begin{document}
\thispagestyle{empty}
\begin{flushright}
KUCP0148\\
March. 21, 2000\\
\end{flushright}
\vskip 2 cm
\begin{center}
{\LARGE \bf Birth of the Brane World }
\vskip 1.7cm

 {\bf Kazuya Koyama}\footnote{
E-mail: kazuya@phys.h.kyoto-u.ac.jp} 
{\bf and}
{\bf Jiro Soda}
\footnote{E-mail: jiro@phys.h.kyoto-u.ac.jp } \

\vskip 1.5mm

\vskip 2cm
 $^1$ Graduate School of Human and Environment Studies, Kyoto University, 
       Kyoto  606-8501, Japan \\
 $^2$ Department of Fundamental Sciences, FIHS, Kyoto University,
       Kyoto, 606-8501, Japan \\
\end{center}

{\centerline{\large\bf Abstract}}
\begin{quotation}
\vskip -0.4cm
Birth of the brane world is studied using the Hamiltonian approach. 
It is shown that an inflating brane world can be created from nothing.
The wave function of the universe obtained from the Wheeler de-Witt equation
and the time-dependent Schr$\ddot{\mbox{o}}$dinger equation for
quantized scalar fields on the brane are the same as in the conventional 
4-dimensional quantum cosmology if the bulk is exactly the Anti-de
Sitter spacetime. The effect of the massive objects in the bulk is also 
discussed. This analysis tells us the presence of the extra dimension 
imprints a nontrivial effect on the quantum cosmology of the
brane world. This fact is important for the analysis of the
quantum fluctuations in the inflationary scenario 
of the brane world.

\end{quotation}
 \newpage

\section{Introduction}

\hspace{1cm}

Much attention has been paid to the possibility we are living
inside a 3-brane in higher dimensional spacetime \cite{Rubakov1,Akama}.
The idea of a brane world was renewed  
by the Horava-Witten theory which relates the strongly coupled 
$E_8 \times E_8$ string heterotic theory to the eleven dimensional 
M-theory compactified on an $S_1/Z_2$ orbifold with a set of $E_8$ 
gauge field at each 10-dimensional fixed plane \cite{Horava}. 
In this theory, the pure supergravity lives in the bulk  
which appears 5-dimensional Anti de-Sitter (AdS) spacetime, 
while the standard model particles are confined to the 3-brane 
\cite{Lukas}. 
Phenomenologically, this picture opens up a route towards resolving the
mass hierarchy between fundamental scales of particle physics and gravity
\cite{Extra,RS1}.
Recently, a simple model based on this picture was constructed by 
Randall and Sundrum \cite{RS2}. 
In their setting, a 4-dimensional domain wall sits 
at a 5-dimensional Ads spacetime. It has been shown the Einstein gravity 
is recovered on the wall with positive tension in the low energy limits
\cite{Shiromizu}. 

This model gives a new setting for the early universe cosmology, which
has been traditionally studied in the framework of the 4-dimensional
action. The cosmological 
evolution of the brane world has been investigated by many 
authors \cite{Cosmology}.  
It has been shown the Friedman equation is recovered
in the low energy limits. In the early universe, the inflationary 
solution plays an important role. 
The solution for the inflating brane world has been obtained  
\cite{Inflation}.

In the traditional 4-dimensional theory, the application of laws of
quantum mechanics to the universe shed light on issues such as the 
initial conditions of the universe. The possibility arises
that the inflating universe can be created from nothing 
by quantum tunneling \cite{Vilenkin}. 
Furthermore, the Wheeler de-Witt (WDW) equation
which describes the evolution of the wave function of the universe
leads to the time-dependent Schr$\ddot{\mbox{o}}$dinger equation 
for quantized matter in the de Sitter spacetime \cite{Rubakov2}. 
This gives the explanation of the origin of the structure of the universe
\cite{Halli}.
Then it seems natural to ask how to implement the quantum 
cosmological idea in the context of the brane world \cite{Sasaki}.  
Can the inflating brane world be created from nothing ?
What is the wave function of the universe ?  How to derive the 
Schr$\ddot{\mbox{o}}$dinger equation for the quantized matter on the brane ?

In this letter, we derive the effective action for the brane world
from the 5-dimensional action and reply these questions. 
From the 5-dimensional viewpoint, the creation of the brane is
the quantum tunneling of the domain wall. The most promising way to deal 
with this process is the Hamiltonian approach 
\cite{Pol, Eardley}. 
We derive the Hamiltonian form of the effective action for the brane 
world and derive the WDW equation. 
We also discuss the effect of the massive objects in the bulk. 
This analysis gives the insight to
the effect of the deviation of the bulk from the AdS spacetime on the 
quantum cosmology of the brane world, which is important for the
analysis of the quantum fluctuations in the inflationary scenario
of the brane world.


\section{The effective action}
In this section we derive the effective action for the brane world.
Our starting point is the 5-dimensional action for the bulk gravity plus 
4-dimensional domain wall; 
\begin{equation}
S= S_G+S_W=\frac{1}{16 \pi G} \int d^5 x \sqrt{-g} ({\cal R}^5 + 2 \Lambda)
- \frac{\mu}{2 \pi^2} 
\int_{wall} d^4 A,
\end{equation}
where ${\cal R}^5$ is the 5-dimensional Ricci scalar, $\Lambda$ is the 
cosmological constant ($\Lambda>0$ for the Anti-de Sitter spacetime),
$G$ is the Newton constant in the 5-dimensional spacetime, 
$\mu/2 \pi^2$ is the energy per unit volume of the domain wall and
$\int_{wall} d^4 A$ is the volume of the wall.
We assume the spacetime is spherically symmetric. The general
spherically symmetric metric can be written as
\begin{equation}
ds^2=-(N^t(r,t) dt)^2+L(r,t)^2(dr + N^r(r,t) dt)^2 + R(r,t)^2 d \Omega_3^2.
\end{equation}
Let us denote the trajectory of the wall in this spacetime as $r_0=r_0(t)$.
Here $r_0$ is the coordinate radius of the wall which describes the
the degree of the freedom of the wall. 
Then the action for the wall becomes 
\begin{equation}
S_W= - \mu \int dt R_0^3 \sqrt{N_0^{t2}-L_0^2(\dot{r}_0+N^r_0)^2}.
\end{equation}
Here the quantities on the wall are denoted like as $R_0=R(r=r_0,t)$.
Defining the conjugate mometa
\begin{eqnarray}
\pi_R &=& \frac{R^2}{N^t} \left(-\frac{2L}{R} \dot{R}-\dot{L}+
(N^r L)'+\frac{2 L}{R} R' N^r \right), \nonumber\\
\pi_L &=& \frac{R^2}{N^t} (N^r R'-\dot{R}), \nonumber\\
p &=& \frac{ \mu R_0^3 L_0^2(\dot{r}_0+N^r_0)}{\sqrt{N_0^{t2}-
L_0^2(\dot{r}_0+N^r_0)^2}},
\end{eqnarray}
the Hamiltonian form of the action is given by
\begin{equation}
S=\int dt \: \left\{
p \: \dot{r}_0+ \int dr \left( \pi_L \dot{L}+ \pi_R \dot{R}-N^t {\cal H}_t
-N^r {\cal H}_r \right)
\right\},
\end{equation}
where
\begin{eqnarray}
{\cal H}_t &=&\frac{4 G}{3 \pi} \left( \frac{\pi_L^2 L}{R^3}
-\frac{\pi_L \pi_R}{R^2} \right) - \frac{3 \pi}{4 G} \left( 
-R^2 \left(\frac{R^{'}}{L} \right)^{'}+LR
\left( 1 -\left( \frac{R^{'}}{L} \right)^2 \right) 
+\frac{\Lambda}{3} L R^3 \right) \nonumber\\
&&+  \delta(r_0-r) \left( \frac{p^2}{L^2} + \mu^2 R_0^6 \right)^{1/2},
\nonumber\\
{\cal H}_r &=& -L \pi_L^{'}+R' \pi_R- \delta(r_0-r) p.
\end{eqnarray}
For a while, we will work in units $4 G/3 \pi=1$.
    
Let the radial coordinate takes the range $r_1 \leq r \leq r_2$. The spacetime
is divided into two regions by the domain wall;
\begin{eqnarray}
V_1: &\quad& r_1 \leq r \leq r_0, \nonumber\\
V_2: &\quad& r_0 \leq r \leq r_2,
\end{eqnarray}
We impose the $Z_2$ symmetry across the wall, then $r_2=2 r_0-r_1$.
The spacetime is obtained by deleting the spacetime region from 
the wall to the boundary and gluing two copies of the remaining
spacetime along the 4-sphere at $r=r_0$. We assume $R(r)$ to be continuous and
$\pi_R$ and $\pi_L$ to be free from $\delta$ functions at the wall. The
integration of the constraints implies the following junction conditions
at the wall;
\begin{equation}
\triangle \pi_L = -\frac{p}{L}, \quad
\triangle R' = - \frac{E}{R^2}, 
\end{equation}
where $E=(p^2+\mu^2 L_0^2 R_0^6)^{1/2}$ is the energy of the wall.
 
We impose the coordinate gauge condition and the slicing condition
\begin{equation}
L=1, \quad R \pi_R=2 \pi_L.
\end{equation}
From these conditions, we obtain the shift function and lapse function as
\begin{eqnarray}
N^r=0,\quad N^t=-\frac{R^2 \dot{R}}{\pi_L}.
\end{eqnarray}
The induced metric on the wall is written as
\begin{equation}
ds^2 \vert_{wall}=- d \tau^2 +R_0^2 d \Omega_3^2,
\end{equation}
where $\tau$ is the propertime of the wall.
The radius at the wall $R_0$ can be identified with the scale
factor of the 3-brane world.

We shall reduce the action to have only the degree of the 
freedom of the brane, that is, $R_0$.  For this purpose, 
we shall solve the constraints in the bulk.
Using the slicing and gauge fixing conditions (9), we can 
solve the momentum constraints and Hamiltonian constraints;
${\cal H}_t=0$ and ${\cal H}_r=0$.
We obtain 
\begin{equation}
 \pi_R =  2 \:  R(r,t) \: \xi_j(t),  \quad
 \pi_L =  R(r,t)^2 \: \xi_j(t), \quad  
  R'= (-1)^\sigma \sqrt{1+ \xi_i(t)^2 
       +\frac{\Lambda R(r,t)^2}{6}-\frac{2 M}{R(r,t)^2}},
\end{equation}
where $\xi_j(t)$ is the constant of integration with respect to the
radial coordinate $r$ in region $V_j$, 
and $\sigma=0$ for $j=1$ and $\sigma=1$ for $j=2$. $M$ is
another constant of integration which is related to the mass
of the 5-dimensional AdS-Schwartzshild black hole.
Making use of these solutions, we shall rewrite 
the gravitational part of the action in the bulk;
\begin{equation}
S_G=\int dt dr \pi_R \dot{R}=\int dt dr 2 \dot{R}\left( 
R \xi_1 \theta(r_0-r) +  R \xi_2 \theta(r-r_0) \right).
\end{equation}
Here we introduced the step function $\theta(x) = 1$ for $x>0$ and 
$\theta(x) =0$ for $x<0$.
Carrying out the integration by parts, this action (13) can be 
rewritten as
\begin{equation}
S_G=\int dt \left\{  \dot{r}_0 
R^2_0 (\xi_2-\xi_1)
+\int dr \left(-   \dot{\xi}_1 R^2 \theta(r_0-r) 
-   \dot{\xi}_2 R^2 \theta(r-r_0) \right)\right\}.
\end{equation}
Next consider the contributions from the constraint;
\begin{equation}
S_C= \int dt dr (-N^t {\cal H}_t
-N^r {\cal H}_r).
\end{equation}
In the bulk, since the constraints are satisfied, there is no contribution
to $S_C$. On the wall, however, since the solutions of the constraint are not 
held, $R''$ and $\pi_L'$ in ${\cal H}_t$ and ${\cal H}_r$ give the
contribution to $S_c$;
\begin{equation}
\int^{r_0+\epsilon}_{r_0-\epsilon} dr 
\left(-N^t(R^2 R'')-N^r(-\pi_L') \right)
=-N^t R^2 \triangle R'+N^r \triangle \pi_L.
\end{equation}
Then we obtain
\begin{equation}
S_{C}=\int dt \left\{-N^t(E+R^2 \triangle R')
+N^r(p+ \triangle \pi_L) \right\}.
\end{equation}
The total action (5) becomes
\begin{equation}
S = \int dt \int dr \left(-  \dot{\xi}_1 R^2 \theta_1 
-  \dot{\xi}_2 R^2 \theta_2 \right) 
+ \int dt \left(-N^t (E+R^2_0 \triangle R') +(\dot{r_0}+N^r )\tilde{p} \right),
\end{equation}
where 
\begin{eqnarray}
\tilde{p} &=&p+\triangle \pi_L, \\
\triangle R' &=& -\sqrt{1+\xi_2^2 +\frac{\Lambda R_0^2}{6}-
\frac{2 M}{R_0^2}}
-\sqrt{1+\xi_1^2 +\frac{\Lambda R_0^2}{6}-\frac{2 M}{R_0^2}}, \\
E &=& (\mu^2 R_0^6+R_0^4(\xi_2 - \xi_1)^2)^{1/2}.
\end{eqnarray}

We shall simplify this action. First, we can drop the term
proportional to $\tilde{p}$ from the junction condition $\tilde{p}=0$.
Next we will simplify the Hamiltonian constraint at
the wall. We redefine the lapse function;
\begin{equation}
N^t \equiv R_0^{-4} \tilde{N}^t (E-R_0^2 \triangle R').
\end{equation}
Then 
\begin{eqnarray}
&& N^t (E+R^2_0 \triangle R')
=\tilde{N}^t R_0^{-4}(E^2-R_0^4 (\triangle R')^2) \nonumber\\
&&=\tilde{N}^t \left[\mu^2 R_0^2 -2 \left\{
1+\xi_1 \xi_2
+\frac{\Lambda R_0^2}{6}-\frac{2 M}{R_0^2}
+\sqrt{1+\xi_1^2 + \frac {\Lambda R_0^2}{6}
- \frac{2 M}{R_0^2}}
\sqrt{1+\xi_2^2 + \frac {\Lambda R_0^2}{6}
- \frac{2 M}{R_0^2}}
\right\}\right].\nonumber\\
\end{eqnarray}
We will take the rest frame of the wall i.e. $p=0$. The 
junction condition implies 
\begin{equation}
\xi_1=\xi_2  \equiv \xi(t).
\end{equation}
The constraint can be written by the variable $\xi$ as
\begin{equation}
{\cal H}^t=\tilde{\mu}^2 R_0^2 +\frac{8 M}{R_0^2}-4(1+\xi^2), \quad 
\tilde{\mu}=\mu \sqrt{1-\lambda},
\end{equation}
where we have introduced the parameter $\lambda$ by
\begin{equation}
\lambda=\frac{2 \Lambda}{3 \mu^2},
\end{equation}               
which reduces $1$ if we take the Randall-Sundrum limit \cite{RS1}. 
Finally we shall consider the kinetic term. Using the variable $\xi$, 
the kinetic term becomes
\begin{eqnarray}
(\mbox{Kinetic term}) &=& \int dt \int^{r_0}_{r_1} dr (- 2 \dot{\xi} R^2 )
= -\int dt \: \dot{\xi} \int^{R_0}_{R_1}\: dR  \frac{2 R^2}{R'}.
\end{eqnarray}
Now we obtain the action describing the dynamics of the brane; 
\begin{equation}
S = \int dt 
\left\{
\: \xi \dot{\chi} -N^t \left( \xi^2+1-\frac{\tilde{\mu}^2}{4}R_0(\chi,\xi)^2
-\frac{2M}{R_0(\chi,\xi)^2} \right)
\right\},
\end{equation}
where $R_0(\chi,\xi)$ is determined by 
\begin{equation}
\chi = \int^{R_0}_{R_1} d R \frac{2 R^2}{R'}
=\int^{R_0}_{R_1} dR \frac{2 R^2}
{\sqrt{1+\xi^2 +\Lambda  R^2 / 6-2 M /R^2}}.
\end{equation}
 

\section{The birth of the brane world}
In this section we assume the bulk is exactly AdS spacetime. 
Then, the effective action (28) reduces to the simple form 
which leads to the conventional Wheeler de Witt(WDW) equation.
Then we show the creation of the inflating brane world from
nothing is possible and derive the wave function of the brane
world.

\subsection{Effective action for $M=0$}
In the effective action (28), 
the canonical variables are $(\chi,\xi)$.
For $M=0$ we can take $r_1=R(r_1)=0$. 
The integration in the expression $\chi$ can be done easily,
then we obtain the effective action
\begin{equation}
S= \int dt 
\left\{
\: \xi \dot{\chi} - N^t \left(
 \xi^2 +1-\frac{\mu^3}{8} \frac{1 -\lambda}{f(\lambda)} \chi \right)
\right\}
,\quad
f(\lambda)=\frac{1}{\lambda}-
\left( \frac{1-\lambda}{\lambda^{3/2}} \right)
\log \frac{1+\sqrt{\lambda}}{\sqrt{1- \lambda}},
\end{equation}
where $\lambda$ is defined in (26) and 
$f(\lambda)$ becomes $1$ as taking the  Randall-Sundrum limit 
$\lambda \to 1$.
Carrying out the canonical transformation
\begin{equation}
\chi=\frac{2}{\mu}f(\lambda) R_0^2, \quad \xi= \frac{\mu}{4 f(\lambda)} 
\frac{P}{R_0},
\end{equation}
we can write the reduced action by $R_0$ and $P$;
\begin{eqnarray}
S &=& \int dt \:
\left \{
P \dot{R}_0- N^t  {\cal H}^t 
\right \}
,\quad
{\cal H}^t = P^2+ \left(\frac{9 \pi^2}{4 G_4^2}\right)
R_0^2 \left( 1-  H^2 R_0^2 \right), \nonumber\\
G_4 &=& \frac{2 G^2 \mu}{3 \pi f(\lambda)},\quad 
H^2= \left(\frac{2 G \tilde{\mu}}{3 \pi}\right)^2 
=\frac{2 G_4}{3 \pi} \mu(1-\lambda)f(\lambda).
\end{eqnarray}
Here we restore the 5-dimensional Newton constant $G$.
This is the Hamiltonian form of the effective action for the brane world.

The constant $G_4$ can be identified with the Newton constant in the
brane world. To see this fact, we solve the $r$ dependence of 
the $R(r,t)$. The solution of $R$ in the bulk is given by
\begin{eqnarray}
 R(r,t) = \left\{ 
\begin{array}{ll}
\displaystyle{
\cosh \theta_1(t) \sqrt{\frac{6}{\Lambda}} \sinh 
\sqrt{\frac{\Lambda}{6}} r}  ,&
\qquad 0< r < r_0,  \\
\\
\displaystyle{
\cosh \theta_2(t) \sqrt{\frac{6}{\Lambda}} \sinh 
\sqrt{\frac{\Lambda}{6}} (2 r_0 - r)},&
\qquad r_0 \leq r < 2 r_0, \end{array}
\right.
\end{eqnarray}
Here, the integration constants are denoted as $\theta_1$ and $\theta_2$.
The location of the wall $r_0$ is determined by the junction condition. 
We will consider on the rest frame of the wall $p=0$. 
The junction conditions (8) become
\begin{equation}
\pi_L(r=r_0 +\epsilon)=\pi_L(r=r_0-\epsilon),\quad
R'(r=r_0 \pm \epsilon)= \mp \frac{1}{2} \mu R_0.
\end{equation}
From these conditions (33), we obtain
\begin{equation}
\theta_1 = \theta_2 = \theta, \quad
\sinh \sqrt{\frac{\Lambda}{6}} r_0
=\sqrt{\frac{\lambda}{1-\lambda}}.
\end{equation}
Thus, the solution is given by $R(r,t)= a(t) \:\: W(r)$, where
\begin{eqnarray}
a(t)&=& H^{-1}\cosh H \tau,\quad \tau=H^{-1} \theta(t),\nonumber\\
W(r)&=& \cosh \sqrt{\frac{\Lambda}{6}}w
- \frac{1}{\sqrt{\lambda}} \sinh \sqrt{\frac{\Lambda}{6}}
\vert w \vert ,\quad w=r-r_0.
\end{eqnarray}
Using this solution the lapse function can be written as
$N^t=\dot{\theta} H^{-1} W(r)$.
Then the classical solution of the spacetime is obtained
\begin{equation}
ds^2 =dw^2+ W^2(w)\gamma_{ij} dx^i dx^j, \quad 
\gamma_{ij} dx^i dx^j=-d \tau^2+ a(\tau)^2 d \Omega_3^2.
\end{equation}
The factor $W(w)$ is often called the warp factor. 
The effective action for 4-dimensional gravity is given by \cite{RS1} 
\begin{equation}
S_{eff} \sim \frac{1}{G} \int d^4x 
\left( 2 \int^{r_0}_0 dw (W(w))^2 \right) \sqrt{\gamma} {\cal R}^{\gamma}
= \frac{1}{G_4} \int d^4x \sqrt{\gamma} {\cal R}^{\gamma}.
\end{equation}
Here ${\cal R}^{\gamma}$ is the Ricci scalar made out of $\gamma_{ij}$.
Hence $G_4$ can be identified with the Newton constant in the brane world.
In the $\lambda \to 1$ limits, the brane world becomes Minkowski spacetime.
In this case, $G_4$ can be written as $G_4= G(\Lambda/6)^{1/2}$
\cite{RS1}. 

\subsection{The birth of the brane world}

Based on the effective action (32), we
discuss the creation of the brane world and it's wave function.
There is no classical solutions for $\lambda \geq 1$. Then
we assume $\lambda <1$. 
The phase space $(R_0,P)$ is described as follows;
\begin{eqnarray}
H^{-1}&<& R_0, \quad \quad \quad \quad \:\:
\mbox{classically allowed region}, \nonumber\\
0 &<&  R_0 < H^{-1}, \quad \mbox{classically forbidden  region}.
\end{eqnarray}
Choosing the lapse function appropriately, the classical equation of
motion is obtained as follows
\begin{equation}
\left(\frac{\dot{R_0}}{R_0}\right)^2=- \frac{1}{R_0^2}+ H^2. 
\end{equation}
The universe can contract from the infinite size and bounce 
at a minimum radius $H^{-1}$ and then re-expand classically. 

We shall quantize this system as a one-dimensional particle system.
We take the wave function of the brane world as $\Psi(R_0)$.
Using the representation
\begin{equation}
P \to -i \frac{d}{d R_0},
\end{equation}
we write the constraint as a Wheler de-Witt (WDW) equation 
${\cal H}_t \Psi(R_0)=0$;
\begin{equation}
\left(- \frac{1}{2} R_0 \frac{d}{d R_0} R_0^{-1} \frac{d}{d R_0}
+ V(R_0) \right) \Psi(R_0)=0 ,\quad
V(R_0)=\frac{1}{2}
\left(\frac{9 \pi^2}{4 G_4^2}\right) R_0^2  
(1- H^2 R_0^2),
\end{equation}
where the choice of the factor ordering is made, which is not
important at the semiclassical level. 
Consider the boundary conditions of this equation (42). In the classical 
forbidden region, the tunneling from $R_0=0$ to $R_0=H^{-1}$ is possible.
From the 5-dimensional point of view, this is a tunneling of the domain wall.
The boundary condition that selects the wave function representing the 
tunneling requires that  $\Psi(R_0)$ should be only an outgoing wave at 
$R_0 \to \infty$. In the classical forbidden region the wave function 
is given by the linear combination of the growing and the decaying solutions.
The wave function corresponds to this boundary condition is given by
\begin{equation}
\Psi(R_0) \propto \mbox{Ai}[z(R_0)]\mbox{Ai}[z(H^{-1})] 
+ i \: \mbox{Bi}[z(R_0)] \mbox{Bi}[z(H^{-1})],
\end{equation}
where Ai and Bi is the Airy functions and $z(R)=(3 \pi/4 G_4
H^2)^{3/2}(1-H^2 R_0^2)$ \cite{Vilenkin}. 
Under-barrier region, the decaying solution dominates. The
semiclassical wave function is given by
\begin{equation}
\Psi(R_0) \sim \exp \left(- \int^{R_0}_0  d R
\sqrt{2 V(R)}  \right).
\end{equation}
Then the tunneling probability is obtained as
\begin{equation}
P \sim \left \vert \frac{\Psi(H^{-1})}{\Psi(0)} \right \vert^2 \sim 
\exp \left(-\frac{\pi}{G_4 H^2} \right).
\end{equation}
From the four dimensional point of view, 
a tunneling from $R_0=0$ to $R_0=H^{-1}$ 
represents the creation of the brane world from nothing.
If we identify $G_4$ as the Newton constant in
our brane world, this tunneling probability agrees with
the one traditionally obtained in the 4-dimensional quantum 
cosmology \cite{Sasaki}. 
The wave function of the brane world coincides 
with the Vilenkin's tunneling wave function. 

We shall include the effect of the matter confined to the wall. 
It is usefull to consider a scalar field as the matter 
confined to the wall. For simplicity, we shall consider a
scalar field $\sqrt{2} \pi \phi$ with potential 
$2 \pi^2 U(\phi)$. We assume there is no back reaction to the 
geometry of the bulk and the junction conditions from this scalar filed, 
that is $U(\phi)/\mu \ll 1$. The 4-dimensional gravity couples to the
matter on the wall with the coupling $G_4$.
The WDW equation including the scalar field on the wall is
given by
\begin{eqnarray}
&& \left(- \frac{1}{2}
R_0 \frac{d}{d R_0} R_0^{-1} \frac{d}{d R_0}+ V(R_0)
- \frac{3 \pi}{2 G_4} H_{\phi}(\phi,R_0)
\right)\Psi(R_0)=0, \nonumber\\ 
&& H_{\phi}= \int \frac{d^3 x}{2 \pi^2} \left(
-\frac{1}{2 R_0^2} \frac{\partial}{\partial \phi^2} + 
\frac{R_0^2}{2}(\partial_i \phi)^2+R_0^4 U(\phi) \right).
\end{eqnarray}
We put the WKB solution of the WDW equation of the form
\begin{equation}
\Psi(R_0,\phi)=C(R_0) e^{-i S(R_0)} \psi(R_0,\phi),\quad
S(R_0)=\int^{R_0}_0  d R \sqrt{- 2 V(R)}.  
\end{equation}
Then we find $\psi(R_0,\phi)$ satisfies the time-dependent 
Schr$\ddot{\mbox{o}}$dinger equation 
\begin{equation}
i \frac{\partial \psi}{\partial \eta}=H_{\phi} \psi,\quad
\frac{d \tau}{d \eta}=R_0,\quad
R_0(\tau)=H^{-1} \cosh H \tau. 
\end{equation}
This is a Schr$\ddot{\mbox{o}}$dinger
equation for a quantized scalar field in the de Sitter spacetime. 

  
\section{Effect of the black hole in the bulk}

\subsection{Effective action for $M \neq 0$}
In this section we discuss the effect of the spherically symmetric
objects in the bulk. The bulk deviates from the AdS spacetime to
the AdS-Schawartzshild spacetimes. 
For $M \neq 0$, there is an event horizon at $r=r_h$ at which 
\begin{equation}
R_h(r_h)=\frac{3}{\Lambda} \left(-1+
\sqrt{1+\frac{4 M \Lambda}{3}}
\right).
\end{equation}
We will take some $r_1$ at which $R(r_1)> R_h$. 
We rewrite the effective action 
by the canonical variable $R_0$. We must use slightly different 
canonical transformation from the previous one. This is related to the fact
$R(r,t)$ cannot be separable into the warp factor and the scale 
factor as is done for $M=0$. 
We use the similar method developed by Kolitch and Eardley \cite{Eardley}. 
By taking the integration by part, the kinetic term can be written as
\begin{eqnarray}
\mbox{(kinetic term)}&=& - \int dt \: \dot{\xi}
\int^{R_0} \: dR \frac{2 R^2}{R'} = - \int dt \: \dot{\theta} \left \{
 \left[ \frac{2 R^3}{3}
\frac{\cosh \theta}{\sqrt{\cosh^2 \theta +\Lambda R^2/6 -2M/R^2}} 
\right ]^{R_0} \right. \nonumber\\
&+& \left.
\int^{R_0} dR \left(\frac{2 R^3}{3} \right) 
\left( \frac{\Lambda R}{6}+\frac{2 M}{R^3} \right)
\frac{\cosh \theta}{ \left(\cosh^2 \theta 
+\Lambda R^2/6-2M/R^2 \right)^{3/2}} \right \}.
\end{eqnarray}
Here we changed the variables as $\xi= \sinh \theta$.
It is convenient to introduce the quantity $T$;
\begin{equation}
T = - \sinh \theta \int^{R_0} dR \left ( \frac{
(2 R^3/3) \:( \Lambda R/6 + 2 M/R^3)}{\sqrt{\cosh^2 \theta 
+ \Lambda R^2/6 -2 M/R^2} (1 + \Lambda R^2/6-2M /R^2)}\right ).
\end{equation}
We can show the total derivative of $T$ with respect to $t$ is given by
\begin{equation}
\dot{T}= \mbox{(kinetic term)}+\frac{2 R_0^3}{3} F(R_0,\theta),
\end{equation}
where
\begin{equation}
F(R_0,\theta)= \dot{\theta} \frac{\cosh \theta}{\sqrt{\cosh^2 \theta
+\Lambda R_0^2/6 -2 M /R_0^2}} - \dot{R_0} \frac{(\Lambda R_0/6
+2 M/R_0^3) \sinh \theta}{\sqrt{\cosh^2 \theta
+\Lambda R_0^2/6 -2 M /R_0^2}(1+ \Lambda R_0^2/6 -2 M/R_0^2)}.
\end{equation}
We introduce the variable $\psi$ by
\begin{equation}
\sinh  \psi = \frac{\sinh \theta}{\sqrt{1+ \Lambda R_0^2/6 -2 M/R_0^2}}.
\end{equation}
The total derivative of the variable is written as
\begin{equation}
 \dot{\psi} = F(R_0,\theta),
\end{equation}
Then defining $\zeta=2 R_0^3/3$, we can write the effective action 
up to the total derivative 
\begin{equation}
S= \int dt \:
\left \{  
\psi  \dot{\zeta} - N^t
\left( \mu^2 \left(\frac{3}{2} \zeta \right)^{2/3}-4
\left( 1+ \frac{\Lambda}{6} \left(\frac{3}{2} \zeta \right)^{2/3}
-2 M \left(\frac{3}{2} \zeta \right)^{-2/3} \right) \cosh^2 \psi \right)
\right \}.
\end{equation}
Now, performing the canonical transformation 
\begin{equation}
\psi=\frac{P}{2 R_0^2},\quad \zeta= \frac{2}{3} R_0^3,
\end{equation}
we rewrite the constraint as
\begin{equation}
P= \pm \left(
\frac{3 \pi}{2 G}\right) R_0^2 \:  \mbox{arcsinh} \sqrt{
\frac{-R_0^2+H^2 R_0^4 +(8 G/3 \pi) M}{R_0^2+(\Lambda/6)R_0^4
-(8 G/3 \pi) M}} .
\end{equation}
There is an ambiguity in the choice of the sign. The choice of 
the sign determines the boundary condition of the wave function.
We will choose the sign so that the wave function represents
the tunneling. After doing this procedure, we obtain
\begin{equation}
S=\int dt \:  
\left \{
 P \dot{R_0}-N^t \left(P - \left(
\frac{3 \pi}{2 G}\right) R_0^2 \:  \mbox{arcsinh} \sqrt{
\frac{-R_0^2+H^2 R_0^4 +(8 G/3 \pi) M}{R_0^2+(\Lambda/6)R_0^4
-(8 G/3 \pi) M}} \right)
\right \},
\end{equation}
where we restore the 5-dimensional Newton constant $G$. 

\subsection{Effect of bulk black hole}
The turning point where $P=0$ is given by
\begin{equation}
R_{\pm}=\frac{1 \pm \sqrt{1- (32G/3 \pi) H^2 M}}{2 H^2}.
\end{equation}
For $ 1 < (32G/3 \pi)H^2 M $ there is no turning point. 
Then we set $0< (32G/3 \pi) H^2 M <1$. 
The phase space is described as follows;
\begin{eqnarray}
R_1 &<& R_0 < R_-, \quad \quad \:\:
\mbox{classically allowed region}, \nonumber\\
R_- &<&  R_0 < R_+, \quad \quad 
\mbox{classically forbidden  region},\nonumber\\
R_+ &<& R_0, \quad \quad \quad \quad \:\:
\mbox{classically allowed region}.
\end{eqnarray}
Classical equation of motion is obtained by 
\begin{equation}
\left(\frac{\dot{R_0}}{R_0}\right)^2=- \frac{1}{R_0^2}+ H^2
+\frac{8 G}{3 \pi} \frac{M}{R_0^4}. 
\end{equation}
The additional term proportional to $M$ behaves as the radiation. 

Taking the momentum as a differential 
operators acting on the wave function of the brane world $\Psi(R_0)$
\begin{equation}
P \to -i \frac{\partial}{\partial R_0}.
\end{equation}
we obtain the WDW equation 
\begin{equation}
\frac{\partial}{\partial R_0} \Psi(R_0)=
-\left(\left(\frac{3 \pi}{2 G}\right) R_0^2 
\arcsin \sqrt{
\frac{R_0^2-H^2 R_0^4 - (8 G/3 \pi) M}{R_0^2+(\Lambda/6)R_0^4
-(8 G/3 \pi) M}}
\right) \Psi(R_0) \equiv - \Sigma(R_0) \Psi(R_0).
\end{equation}
The tunneling rate at the semiclassical order can be evaluated by
\begin{equation}
P \sim \exp \left(-2 \int^{R_+}_{R_-} d R \: \Sigma(R) \right).
\end{equation}
We can verify this agrees with the result obtained in the previous section
for the limit $M \to 0$. For $M \neq 0$,
the inflating brane world is created by the tunneling from the 
closed FRW universe driven by the massive objects in the bulk.
For a small radius of the brane world $R_0 < (8 G M/3 \pi H^2)^{1/4}$, 
the wave function is different from the one
obtained in 4-dimensional quantum cosmology. This implies 
the presence of the extra dimension gives the nontrivial effect on the
quantum cosmology of the 4-dimensional universe 
if the bulk deviates from the AdS spacetime. 
Technically, this comes from the
fact the $r$ dependence in $R(r,t)$ cannot be separable. For large
$R_0$, the effect of the massive objects in the bulk on the evolution
of the brane world becomes negligible, then the wave function of the
universe becomes the familiar one obtained in the previous section  
for $M=0$.


\section{Summary and Discussion}
\hspace{1cm}\\

In this letter, we derived the effective action for the brane world
from the 5-dimensional action for the bulk gravity plus 4-dimensional
 domain wall. Based on this action, we derived the WDW equation for 
the brane world.
If the bulk is AdS spacetime, the brane world can be created from nothing.
The WDW equation and the time-dependent
Schr$\ddot{\mbox{o}}$dinger equation for the 
quantized matter on the inflating brane are the same as those in
the conventional 4-dimensional quantum cosmology.
It is easy to extend our analysis to the many brane case. In that case,
the ``time'' is defined in each brane world as is suggested by the
derivation of the Schr$\ddot{\mbox{o}}$dinger equation.

We discussed the effect of the spherically symmetric massive
objects in the bulk. For a small radius of the brane world, 
the bulk deviates from the AdS spacetime largely. 
In this case we found the presence 
of the extra dimension affects the quantum cosmology of the brane world, 
then the wave function is different from the one 
obtained in 4-dimensional quantum cosmology. 

The most important success of the inflationary scenario is that 
this gives the origin of the structure of the universe. The analysis
of the evolution of the quantum fluctuations is needed to explore this 
issue in the context of the brane world.
We must include the effect of the back reaction to the geometry of the bulk
in the Schr$\ddot{\mbox{o}}$dinger equation for the quantized matter on
the wall. It has been shown the deviation 
of the bulk spcaetime from the exact AdS spacetime is essential to
describe the inhomogeneous brane world \cite{Shiromizu}. 
The analysis for the bulk with black holes tells us if the bulk
deviates from the Ads spacetime, the presence of the extra dimension imprints
a non-trivial effects on the quantum cosmology of the brane world.
A detailed investigation of the evolution of the quantum fluctuations 
will be left for the future work \cite{KK}.


\section*{Acknowledgements}
The work of J.S. was supported by Monbusho Grant-in-Aid No.10740118
and the work of K.K. was supported by JSPS Research Fellowships for
Young Scientist No.4687. 
K.K also thanks Kayoko Maeda for useful discussions.


\end{document}